\documentclass[letter,longauth]{aa} 
\usepackage{graphicx}
\usepackage{txfonts}
\usepackage{color}
\newcommand{\um}{$\mu$m}                                 
\newcommand{\rsun}{$R_{\odot}$}

\newcommand{\acena}{$\alpha \, {\rm Cen\,A}$}          
\newcommand{\acenb}{$\alpha \, {\rm Cen\,B}$}                          %
\newcommand{\acen}{$\alpha \, {\rm Cen}$}                              %
\newcommand{\lapprox}{$\stackrel {<}{_{\sim}}$}
\newcommand{\about}{$\sim$}                       
\newcommand{\asec}{$^{\prime \prime}$}
\newcommand{\adeg}{$^{\circ}$}

\newcommand{\adegdot}[2]{\mbox{#1$\stackrel {\circ}{_{\bf \cdot}}$#2}}

\newcommand{\asecdot}[2]{\mbox{#1$\stackrel {\prime \prime}{_{\bf \cdot}}$#2}}

\begin{document}

   \title{ALMA observations of  $\alpha$ Centauri}

   \subtitle{First detection of main-sequence stars at 3\,mm wavelength}

   \author{
                		R. Liseau\inst{1}
          \and
                		W. Vlemmings\inst{1}
          \and 
                		A. Bayo\inst{2}  
         \and
                		E. Bertone\inst{3}
        \and
                        	J.H. Black\inst{1}
        \and
                        	C. del Burgo\inst{3}     
         \and
                        	M. Chavez \inst{3}
         \and
                        	W. Danchi\inst{4}                
         \and        	    	 
		       V. De la Luz\inst{5}             
         \and          
                         C. Eiroa\inst{6} 
         \and
                        S. Ertel\inst{7} 
         \and
                        M.C.W. Fridlund\inst{8}
         \and
                        K. Justtanont\inst{1}
         \and
                        A. Krivov\inst{9}
         \and
                        J.P. Marshall\inst{10} 
         \and
                        A. Mora\inst{11} 
         \and
                        B. Montesinos\inst{12}
         \and
                        L.-A. Nyman\inst{13}
         \and
                        G. Olofsson\inst{14}            
         \and
                        J. Sanz-Forcada\inst{12}
         \and
                        P. Th\'ebault\inst{15} 
          \and
                        G.J. White\inst{16,\,17} 
          }

   \institute{Department of Earth and Space Sciences, Chalmers University of Technology, Onsala Space Observatory, SE-439 92 Onsala, Sweden,
                                 \email{rene.liseau@chalmers.se}     
     \and
                Instituto de F\'{\i}sica y Astronom\'{i}a, Universidad de Valpara\'{\i}so, Av. Gran Breta\~na 1111, Valpara\'{i}so, Chile                                               
     \and
                Instituto Nacional de Astrof\'{\i}sica, \'{O}ptica y Electr\'{o}nica (INAOE), Luis Enrique Erro 1, Sta. Mar\'{i}a Tonantzintla, Puebla, Mexico,             
    \and                             
                Goddard Space Flight Center, 8800 Greenbelt Rd, Greenbelt, MD 20771, United States          
     \and       
                SCiESMEX, Instituto de Geofisica, Unidad Michoacan, Universidad Nacional Autonoma de Mexico, Morelia, Michoacan, Mexico. CP 58190  
    \and                
                Departamento de F\'{i}sica Te\'{o}rica, C-XI, Facultad de Ciencias, Universidad Aut\'{o}noma de Madrid, Cantoblanco, 28049 Madrid, Spain           
     \and       
                European Southern Observatory, Alonso de Cordova 3107, Vitacura, Casilla 19001, Santiago, Chile       
     \and
                Leiden University, Rapenburg 70, 2311 EZ Leiden, Netherlands     
     \and
                Astrophysikalisches Institut und Universit\"atssternwarte, Friedrich-Schiller-Universit\"at Jena, Schillerg\"a\ss chen 2-3, 07745 Jena, Germany        
    \and        
                University of New South Wales, School of Physics, Australia 
    \and
                ESA-ESAC Gaia SOC, P.O. Box 78, 28691 Villanueva de la Ca\~nada, Madrid, Spain 
    \and
                Departamento de Astrof\'{\i}sica, Centro de Astrobiolog\'{\i}a (CAB, CSIC-INTA), Apartado 78, 28691 Villanueva de la Ca\~nada, Madrid, Spain             
   \and
                Joint ALMA Observatory (JAO), Alonso de Cordova 3107, Vitacura, Santiago, Chile                 
   \and
                Stockholm University, AlbaNova University Center, Department of Astronomy, SE-114 19 Stockholm, Sweden          
   \and
                LESIA-Observatoire de Paris, UPMC Univ. Paris 06, Univ. Paris-Diderot, France  
   \and
                Astrophysics Group, Department of Physics \& Astronomy, The Open University, Walton Hall, Milton Keynes MK7 6AA, UK  
   \and        
                Space Science \& Technology Department, CCLRC Rutherford Appleton Laboratory, Chilton, Didcot, Oxfordshire OX11 0QX, UK 
                }
   \date{Received ; accepted}
  \abstract
   {The precise mechanisms that provide the non-radiative energy for heating the chromosphere and the corona of the Sun and those of other stars constitute an active field of research. By studying stellar chromospheres one aims at identifying the relevant physical processes. Defining the permittable extent of the parameter space can also serve as a template for the {\it \emph{Sun-as-a-star}}. This feedback will probably also help identify stars that potentially host planetary systems that are reminiscent of our own.}
   {Earlier observations with {\it Herschel} and APEX have revealed the temperature minimum of \acen, but these were unable to spatially resolve the binary into individual components. With the data reported in this {\it Letter}, we aim at remedying this shortcoming. Furthermore, these earlier data were limited to the wavelength region between 100 and 870\,\um. In the present context, we intend to extend the spectral mapping (SED) to longer wavelengths, where the contrast between stellar photospheric and chromospheric emission becomes increasingly evident.}
   {The Atacama Large Millimeter/submillimeter Array (ALMA) is particularly suited to point sources, such as unresolved stars. ALMA provides the means to achieve our objectives with both its high sensitivity of the collecting area for the detection of weak signals and the high spatial resolving power of its adaptable interferometer for imaging close multiple stars.}
   {This is the first detection of main-sequence stars at a wavelength of 3\,mm. Furthermore, the individual components of the binary \acen AB are clearly detected and spatially well resolved at all ALMA wavelengths. The high signal-to-noise ratios of these data permit accurate determination of their relative flux ratios, i.e., $S_{\!\nu}^{\rm B}/S_{\!\nu}^{\rm A}=0.54\pm0.04$ at 440\,\um, $=0.46\pm0.01$ at 870\,\um, and $=0.47\pm0.006$ at 3.1\,mm, respectively.}
   {The previously obtained flux ratio of $0.44\pm0.18$, which was based on measurements in the optical and at 70\,\um, is consistent with the present ALMA results, albeit with a large error bar. The observed 3.1\,mm emission greatly exceeds what is predicted from the stellar photospheres, and undoubtedly arises predominantly as free-free emission in the ionized chromospheric plasmas of both stars. Given the distinct difference in their cyclic activity, the similarity of their submm SEDs appears surprising. }
   \keywords{stars: chromospheres --  stars: solar-type -- (stars:) binaries: general -- stars: individual: \acen tauri AB -- submillimeter: stars -- radio continuum: stars
               }
   \maketitle
%

\section{Introduction}

About two thirds of the 133 FGK stars observed at wavelengths of 100\,\um\ and 160\,\um\ by {\it Herschel} as part of the DUNES\footnote{DUNES stands for DUst in NEarby Stars, a  {\it Herschel} Open Time Key Program, PI C.\,Eiroa.} program have suggested that temperature minima are present in their atmospheres \citep{eiroa2013}. These stars are commonly recognized as solar-type and the chromospheres and coronae were expected on the basis of  their known CaII\,H\&K indices and X-ray luminosities. 

In the particular case of the nearby binary \acena B (G2\,V and K1\,V), the far-infrared spectral evidence for the phenomenon of a temperature minimum was quite convincing, although the pair was spatially unresolved by our long-wavelength observations \citep{liseau2013,wiegert2014}. Furthermore, since adequate theoretical stellar model atmospheres longward of 40\,\um\ were lacking, relative flux ratios had to be deduced by extrapolation from data obtained at shorter wavelengths. This could potentially lead to large errors, since \acenb\ is known to be considerably more active than the primary \acena\  \citep{ayres2014}, and one might therefore a priori not expect the scaling in the optical (photospheric flux) to also apply to the far-infrared and submillimeter (FIR/sub-mm) regimes, where the radiation originates at higher atmospheric levels in the chromosphere. There, the continuum opacity is controlled by free-free H$^-$ processes and temperatures in these optically thick layers, which follows the photospheric (negative) temperature gradient and samples the temperature inversion in the lower chromosphere (temperature minimum). Thereafter, temperatures start to increase again.

\begin{figure}
  \resizebox{\hsize}{!}{
    \rotatebox{00}{\includegraphics{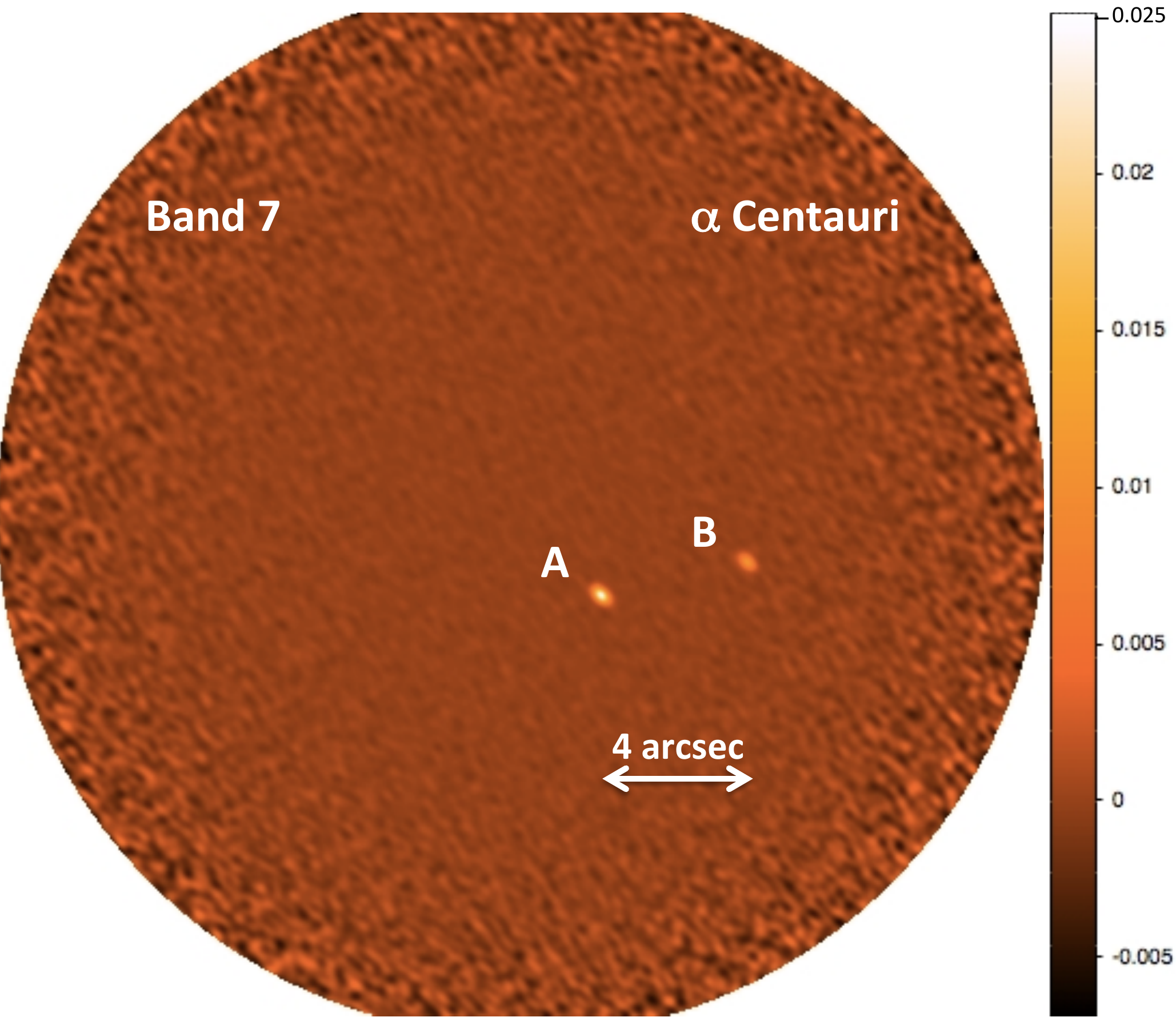}}
                        }
  \caption{ALMA observation of \acena\ and B at 870\,\um\  (\about\,344\,GHz) on 7 July 2014. At 1.35\,pc distance, this binary is nearest to the Sun. The primary has spectral type G2\,V, and the secondary is a K1\,V dwarf. The stellar disks are unresolved, so that the synthesized beam is defined by the point-like stars and beam sizes, are given in Sect.\,2. Intensity units are Jy\,beam$^{-1}$. North is up and east to the left.
   }
  \label{band7}
\end{figure}

The {\it Herschel} photometry permitted the rough spectral coverage of the temperature minimum up to about 500\,\um, where the measurements at the longest wavelengths were at the very limit of the detection capability of the SPIRE instrument \citep{griffin2010}. In addition, the spatial resolution of the 3.5\,m {\it Herschel} single-dish telescope \citep{pilbratt2010} was inadequate for resolving the binary into its components. Both of these shortcomings could be mitigated using the Atacama Large Millimeter/submillimeter Array (ALMA), and in this {\it Letter} we report our initial results of these observations. The ALMA observations and data reduction are described in Sect.\,2, the results are presented in Sect.\,3, and their
relevance discussed in Sect.\,4. Finally, Sect.\,5 briefly lists our main conclusions.

\section{Observations and data reduction}

The binary \acena B was observed in three ALMA continuum bands, with $7.5$\,GHz effective bandwidth spread over four spectral windows in each of the bands.  The observations in Band\,3, centered on 97.5\,GHz (3.1\,mm), were taken on 2014 July 3, and they used 30 antennas and a total observing time of $\sim50$\,min, of which $\sim30$\,min were spent on-source. The Band\,7 observations, centered on 343.5\,GHz (872.8\,\um), were taken on 2014 July 7 using 32 antennas with $\sim20$\,min of observing time with $\sim2$\,min on-source. Finally, the observations in Band\,9, centered on 679\,GHz (441.5\,\um), were taken on 2014 July 18 using 31 antennas and $\sim30$\,min total and $\sim8$\,min on-source observing time. 

\begin{table*}
\caption{ALMA data for \acen}             
\label{data}      
\centering          
\begin{tabular}{l l  l l l cc}     
\hline\hline \\ 
\smallskip
\smallskip     
                         & \multicolumn{4}{c}{Flux density, $S_{\!\nu}$ (mJy)}  &  \multicolumn{2}{c}{$\Delta \log S_{\!\nu}/\Delta \log \nu$} \\
                  \cline{2-4}  \cline{6-7}
\\
                 & \,\,\,\,\,\,\,\,\,\,\,\,\,\,\,\,\,\,\,\,\,\,679\,GHz   & \,\,\,\,\,\,\,\,\,\,\,\,\,\,\,\,\,\,\,\,343.5\,GHz    & \,\,\,\,\,\,\,\,\,\,\,\,\,\,\,\,\,\,\,\,\,97.5\,GHz  &      & &     \\
 Star            & Band 9 (441.5\,\um)  & Band 7 (872.8\,\um)   & Band 3 (3075\,\um)  &  & $\alpha_{9,\,7}$      & $\alpha_{7,\,3}$      \\         
                 & 18 July, 2014 [S/N]  & 7 July, 2014 [S/N]            & 3 July, 2014 [S/N]         &    &                               &                               \\
\hline \\
\acena  & $107.2\pm 1.5$ \,\,\,\,\,[71] & $26.06\pm 0.19$ [137] & $3.373\pm 0.011$ [307]   &        & 2.08  & 1.62  \\
\acenb  & \phantom{1}$57.6\pm 4.5$ \,\,\,\,\,[13] & $12.04\pm 0.23$ \,\,\,[52]& $1.585\pm 0.016$ \,\,\,[99]&& 2.30      & 1.61  \\
\hline                
\end{tabular}
\end{table*}

Calibration was performed using the CASA package\footnote{ CASA is an acronym for {\it Common Astronomy Software Application.}}  following standard procedures and using, for all data sets, the quasar J$1617-5848$ as complex gain calibrator, which was within 12\adeg\ of \acen. The quasar J$1427-4206$, at \adegdot{18}{5} separation,  served as bandpass calibrator. The secondary gain calibrator J$1508-4953$ was observed in Band\,9.

Flux calibration was done using Ceres in Bands\,3 (97.5\,GHz, 3075\,\um) and 9 (679\,GHz, 441.5\,\um), when at 57\adeg\ elevation, while \acen\ was at 50\adeg. Titan was used for Band\,7 (343.5\,GHz, 872.8\,\um), when it was at 59\adeg\ elevation and \acen\ at 43\adeg. Bootstrapping the fluxes of our calibrators yielded a flux for  J$1617-5848$ of $1.555\pm0.002$\,Jy (Band\,3), $0.520\pm0.001$\,Jy (Band\,7), and $0.321\pm0.003$ \,Jy (Band\,9). For J$1427-4206$, we found $2.434\pm0.004$\,Jy and $1.498\pm0.005$\,Jy in Bands 7 and 9, respectively, and for the secondary gain calibrator J$1508-4953$ in Band\,9, we found $0.71\pm0.03$\,Jy.  Based on the flux values for the calibrators that are provided by the observatory, we estimate the {\it \emph{absolute}} flux calibration to be accurate to within $5\%$, $7\%$, and $15\%$ for Bands\,3, 7, and 9, respectively. These calibration data are updated on a regular basis, within at most three weeks, and no flux variations for the calibration sources that exceeded the quoted errors were noticed.

Finally, imaging was performed using Briggs weighting in Band\,7 and natural weighting on Bands\,3 and 9. In Band\,9, \acena\ was strong enough to perform self-calibration, improving the rms noise. In the final, primary beam-corrected images, the rms noise per synthesized beam in the pointing center was $0.02$\,mJy\,beam$^{-1}$ (\asecdot{1}{72}\,$\times$\,\asecdot{1}{50}, at PA = 19\adeg) in Band\,3 (3.1\,mm), $0.2$\,mJy\,beam$^{-1}$ (\asecdot{0}{42}\,$\times$\,\asecdot{0}{28}, 47\adeg) in Band 7 (872.8\,\um), and $1.0$\,mJy\,beam$^{-1}$ (\asecdot{0}{22}\,$\times$\,\asecdot{0}{16}, 35\adeg) in Band\,9 (441.5\,\um). Because an error in the pointing position meant that the binary was offset from the pointing center, the rms toward \acenb\ is somewhat increased as is noticeable in Table\,\ref{data}.

\begin{figure}
  \resizebox{\hsize}{!}{
    \rotatebox{00}{\includegraphics{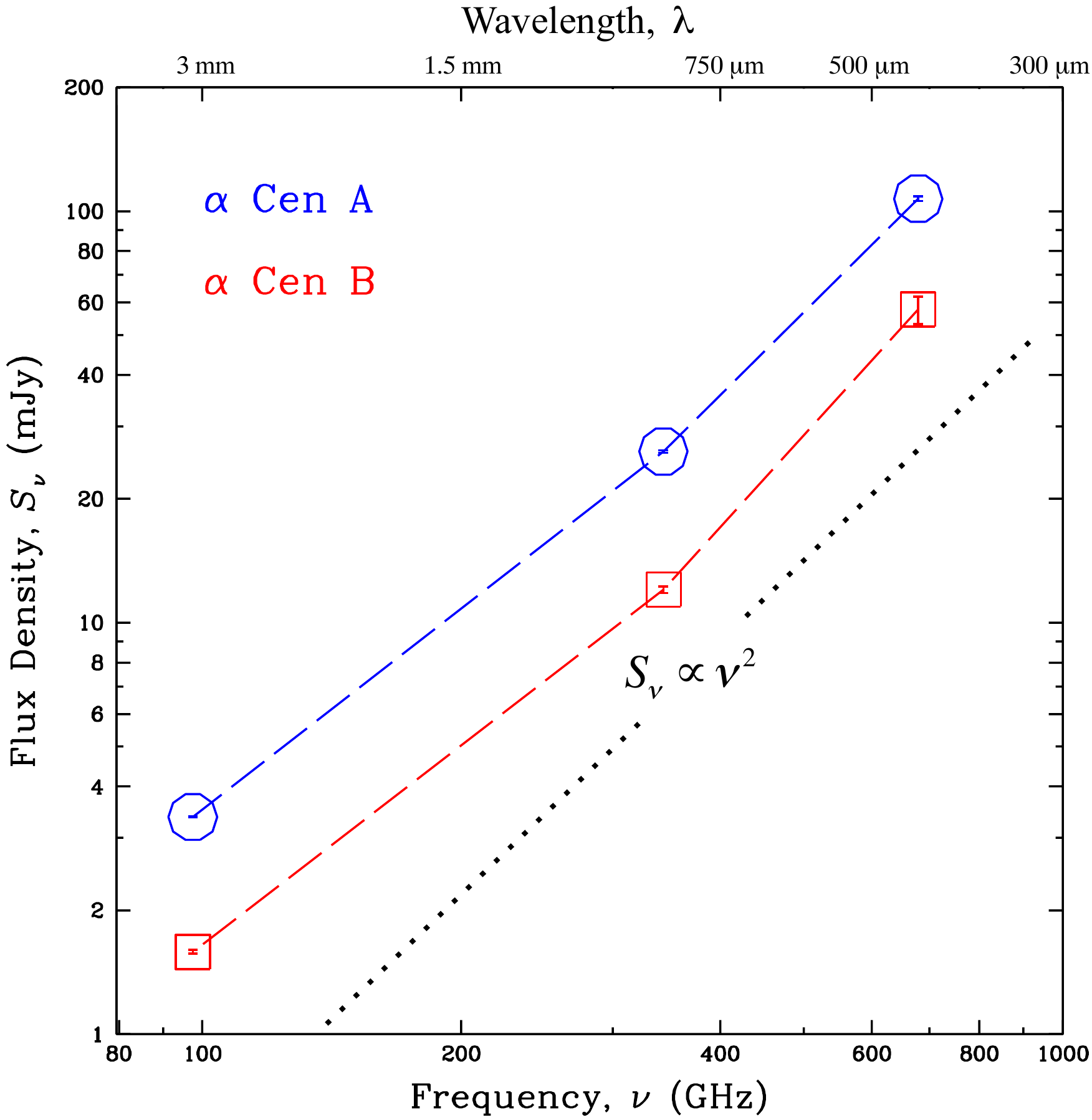}}
                        }
  \caption{ALMA measurements of the flux density of \acena\ (blue circles) and of \acenb\ (red squares), with error bars inside the symbols. For comparison, the black dotted line shows the run of optically thick free-free emission in the Rayleigh-Jeans regime, i.e., $S_{\nu} \propto \nu^2$ (see the text).
     }
  \label{freq}
\end{figure}

\section{Results}

In all bands the binary was clearly detected at high signal-to-noise ratio (S/N) and was well resolved into its individual components. An example is shown in Fig.\,\ref{band7} for the ALMA observation at 872.8\,\um\ (343.5\,GHz, Band\,7).
The measured flux densities are reported in Table\,\ref{data} and also displayed graphically in Fig.\,\ref{freq}. 

The decomposition of the relative contributions to the fluxes of the unresolved binary components in the far infrared was previously based on the comparison of observed photometric measurements with theoretical model atmospheres in the optical and infrared spectral regions. In addition, one observation at 70\,\um\ with {\it Herschel}-PACS \citep{molinari2010,wiegert2014} nearly resolved the binary, which permitted individual fluxes to be measured. Over the wavelength range $0.09 - 70$\,\um,\ the ratio was $S_{\!\nu}^{\rm B}/S_{\!\nu}^{\rm A}=0.44\pm0.18$, which was then also applied to the analysis of the temperature minima of the stars. 

\begin{figure*}
  \resizebox{\hsize}{!}{
    \rotatebox{00}{\includegraphics{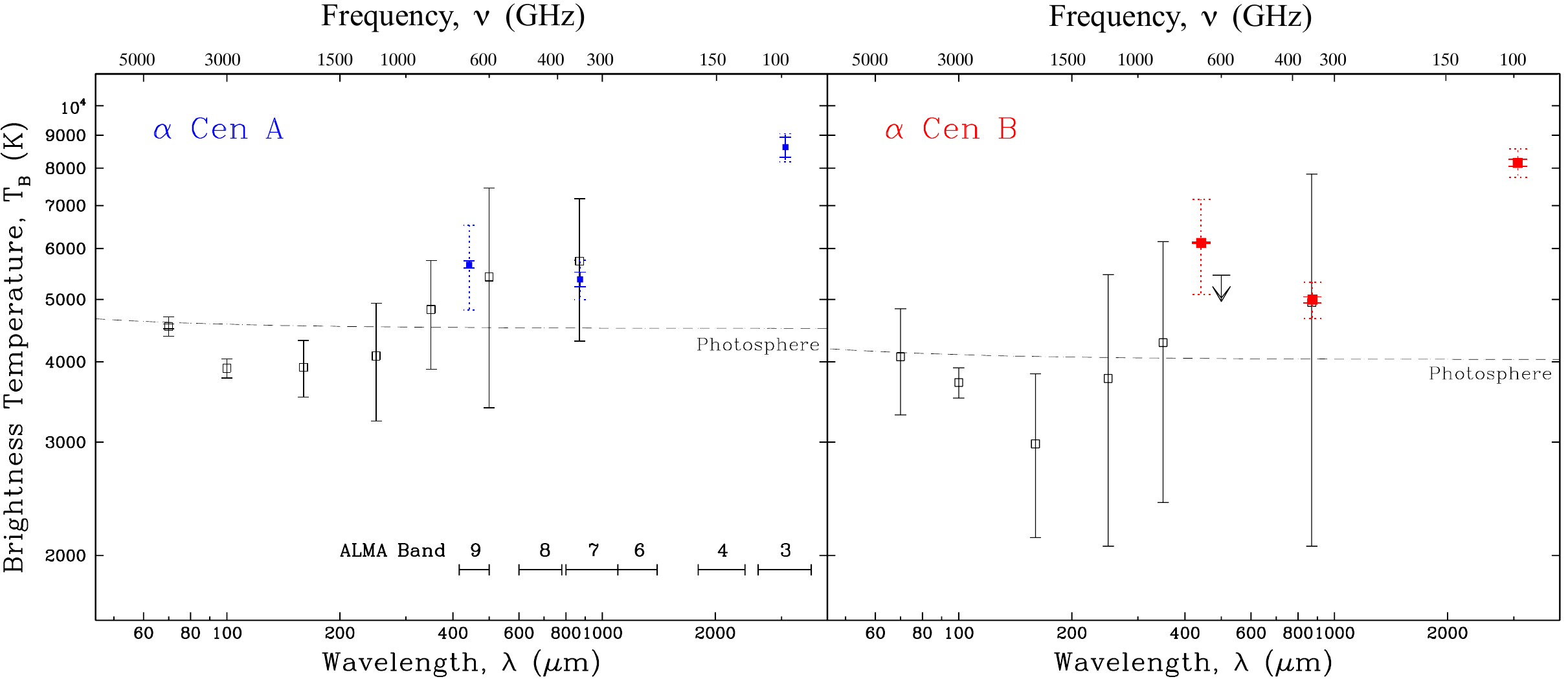}}
                        }
  \caption{Far-infrared to sub-mm/mm SEDs of \acena\ (left) and \acenb\ (right), where blue and red symbols identify the respective ALMA measurements. Observational rms-errors are given by solid bars and maximum {\it \emph{absolute}} uncertainties of the ALMA data as dots, where relative values are $\le 15\%$ for band\,9, $\le 7\%$ for band\,7, and $\le 5\%$ for band\,3. With the exception of the point at 70\,\um, the data shown by black symbols are based on unresolved single-dish observations (see the text). The extrapolations longward of 45\,\um\ of the PHOENIX model atmospheres \citep{brott2005} are indicated by the dashed lines.
   }
  \label{alma_ab}
\end{figure*}

With ALMA, the high S/Ns of the resolved data allow accurate determination of their flux ratios, i.e., $S_{\!\nu}^{\rm B}/S_{\!\nu}^{\rm A}=0.54\pm0.04$ at 440\,\um, $=0.46\pm0.01$ at 870\,\um, and $=0.47\pm0.006$ at 3.1\,mm. In Fig.\,\ref{alma_ab}, the SEDs are presented for both stars in terms of brightness temperature\footnote{$T_{\rm
B}(\nu) = \frac { 2\,\pi\,\hbar\,\nu }{ k } \left [  \ln \left ( \frac{4\,\pi^2\,(R_{\rm
star}+h)^2\,\hbar\,\nu^3}{D^2\,c^2\,S\!_{\nu}} + 1\right ) \right ]^{-1}$
\citep[see, e.g.,][]{liseau2013}}  versus wavelength. Also shown are the levels of the respective photospheres, revealing both the deficiencies ($70\!<\!\lambda$\,\lapprox\,400\,\um) and excesses ($\lambda\!> \!400$\,\um) of the observed emission. Especially at the longest wavelength, i.e. at 3.1\,mm, the chromospheres of both stars are clearly detected well above their photospheric values (and at high S/N, $\ge 100$). 

Leaving aside the Sun for a moment, this ALMA measurement constitutes the first detection of main-sequence stars at 3.1\,mm.

\section{Discussion}

In the Rayleigh-Jeans regime (RJ), optically thick free-free emission (or {\it \emph{Bremsstrahlung}}) will behave as $S_{\!\nu} \propto \nu^2$, so that the  spectral index, $\alpha = \Delta \log S_{\!\nu}/\Delta \log{\nu} = 2$.  For optically thin emission, $\alpha_{\rm {ff}}$ is distinctly different and slightly negative, i.e., essentially zero \citep[see, e.g.,][]{wright1975}.  In Fig.\,\ref{freq}, we present the flux densities $S_{\!\nu}$ of \acena B that were measured with ALMA at three frequencies, and in Table\,\ref{data} we also present the observed values of $\alpha$. In the figure, the dashes connecting the data points are meant to guide the eye and can be compared to the dotted line that shows the spectral shape of optically thick free-free emission. Deviations may be introduced by different depth views at different frequencies along the temperature gradient (down to $\tau_{\nu} \sim 1$), since the dependence of  $S_{\!\nu}$ on $T$ is linear in the optically thick RJ-regime. The measurements seem to indicate changing opacities. However,  a more conclusive statement would require a tighter spectral sampling, including the intermediate ALMA bands.

One can use observations on either side of the turnover frequency $\nu_1$, i.e., where the free-free spectrum turns from optically thick to thin, to construct an empirical model chromosphere.
For \acena\ and B, empirical chromosphere models based on optical line data (Ca\,II K) have been presented by \citet{ayres1976}, with chromospheric temperatures not exceeding 5700\,K.

In the optically thick case, the measured brightness temperature, $T_{\rm B}$, corresponds to the actual temperature of the emitting gas, and once the temperature is known, the density can be derived from the optically thin observations, since $S_{\!\nu} \propto EM\times T^{-1/2}$, and where $EM = \int\!n_{\rm e}\,n_{\rm i}\,dh$ is the emission measure. In the Sun, the height of the chromosphere $\int\!dh$ corresponds to about 0.1\% of its radius. It is reasonable to assume that this will not be very different in the \acen\ stars\footnote{For the stellar radii, we use the data compiled in Table\,4 of \citet{wiegert2014}, i.e. $R_{\rm A}=1.224\pm0.003$\,\rsun\ and $R_{\rm B}=0.863\pm0.003$\rsun.}.
Models of the chromospheres of these stars will be presented in a forthcoming paper.

To obtain data in the optically thin regime would require observations of the stars at longer wavelengths (lower frequencies). However, to the best of our knowledge, such data are not available for \acena\ and B. Observations of this kind have, however, recently been reported for three other solar-type main-sequence stars by \citet{villadsen2014}. That demonstrates that this type of observation has now become feasible with large telescope arrays operating at microwave wavelengths, and they should also be attempted for the \acen tauri binary. With declinations south of $-60$\adeg, the Australia Telescope Compact Array (ATCA) could be an option. 

Observations and/or models of the {\it \emph{quiet}} Sun have been published by, for example, \citet{vernazza1981,loukitcheva2004, fontenla2007,avrett2008}, and \citet{delaluz2014}. Evidence is, however, mounting that the heating of the solar chromosphere and corona is dominated by {\it \emph{active}}, magnetically controlled, processes \citep{carlsson1995}, and it remains to be seen what  {\it \emph{quiet}} stellar chromosphere models could add to the more complete understanding of the physics of the Sun and other stars. However, the Sun is considered to be a relatively inactive star and in this context, it could also be interesting to note that the \acen\ stars show very different levels of activity, with \acena\ being much quieter even than the Sun, whereas \acenb\ is considerably more active \citep[e.g.,][]{ayres2014}. The great similarity in their submm SEDs is therefore quite astounding. Finding the solution to this enigma may also provide valuable feedback for understanding the {\it \emph{Sun-as-a-star}}.

\section{Conclusions}

The nearby solar-type star \acen\ has been clearly detected by ALMA at three wavelengths between 0.4 to 3.1 millimeters.  In particular, the spectacular detection of both stars at 3.1\,mm is the first ever of its kind.
Also, these observations conform to our earlier data, which was obtained by the DUNES team with {\it Herschel}, and confirm the existence of the temperature minima in the atmospheres of the binary components \acena\ and B.

Below, we briefly summarize our main conclusions:

\begin{itemize}

\item[$\bullet$]  At FIR/submm wavelengths, the binary was not spatially resolved by the {\it Herschel}-telescope;  however, the 4\asec\ binary separation in 2014 is easily resolved by the present ALMA observations.

\item[$\bullet$]  The new ALMA data have provided accurate flux ratios that agree with our previously estimated value. 

\item[$\bullet$]  The measured spectral indices are consistent with emission that is dominated by optically thick free-free processes (Bremsstrahlung).

\item[$\bullet$]  The long-wave emission (ALMA-band 3 at 97.5\,GHz) is undoubtedly of chromospheric origin, and originates in material at temperatures in excess of 8000\,K.

\item[$\bullet$]  The chromospheres of the \acen\ stars seem similar in appearance to empirical models of the {\it \emph{quiet}} Sun. However, the ALMA data hint at the possibility that the less active primary \acena, and the more active companion B, both heat their chromospheres to higher temperatures.

\end{itemize}

\begin{acknowledgements}
We thank the referee for a thoughtful and detailed report. Thanks also go to the members of the Nordic ARC node (\texttt{http://www.nordic-alma.se/}), and we wish to thank the ALMA staff for their assistance with the observations. This paper makes use of the following ALMA data: ADS/JAO.ALMA\#2013.1.00170.S. ALMA is a partnership of ESO (representing its member states), NSF (USA), and NINS (Japan), together with NRC (Canada) and NSC and ASIAA (Taiwan), in cooperation with the Republic of Chile. The Joint ALMA Observatory is operated by ESO, AUI/NRAO, and NAOJ.
\end{acknowledgements}

\bibliographystyle{aa}
\bibliography{25189_rl.bbl}

\end{document}